\documentclass{article}

\usepackage{pifont}
\usepackage{spconf,amsmath,graphicx}
\usepackage{bm}
\usepackage{multicol}
\usepackage{multirow}

\usepackage{enumitem}
\usepackage{url,hyperref,cleveref}
\usepackage{diagbox} 
\usepackage{booktabs} 

\usepackage{subfigure}

\title{Contrastive Learning with Audio Discrimination for Customizable Keyword Spotting in Continuous Speech} 
\name{Yu Xi$^1$, Baochen Yang$^1$, Hao Li$^2$, Jiaqi Guo$^2$, $^{\dagger}$Kai Yu$^1$
\thanks{$^{\dagger}$Kai Yu is the corresponding author.}}
\address{$^1$MoE Key Lab of Artificial Intelligence, AI Institute, X-LANCE Lab, Shanghai Jiao Tong University
\\$^2$AISpeech Ltd, Suzhou China}

\begin{document}
\ninept
\maketitle

\linespread{0.8}
\begin{abstract}
Customizable keyword spotting (KWS) in continuous speech has attracted increasing attention due to its real-world application potential. While contrastive learning (CL) has been widely used to extract keyword representations, previous CL approaches all operate on pre-segmented isolated words and employ only audio-text representations matching strategy. However, for KWS in continuous speech, co-articulation and streaming word segmentation can easily yield similar audio patterns for different texts, which may consequently trigger false alarms. To address this issue, we propose a novel CL with Audio Discrimination (CLAD) approach to learning keyword representation with both audio-text matching and audio-audio discrimination ability. Here, an InfoNCE loss considering both audio-audio and audio-text CL data pairs is employed for each sliding window during training. Evaluations on the open-source LibriPhrase dataset show that the use of sliding-window level InfoNCE loss yields comparable performance compared to previous CL approaches. Furthermore, experiments on the continuous speech dataset LibriSpeech demonstrate that, by incorporating audio discrimination, CLAD achieves significant performance gain over CL without audio discrimination. Meanwhile, compared to two-stage KWS approaches, the end-to-end KWS with CLAD achieves not only better performance, but also significant speed-up. 
\end{abstract}

%


\section{Introduction}
Keyword spotting (KWS) technology, which provides an essential entrance to human-machine interaction systems, has been widely used in smart home devices and intelligent cockpits. Although fixed wake word detection systems have reached satisfactory performance, building a user-defined or customizable system is still a considerable challenge, especially in the scenario of KWS in continuous speech. 

There are two main categories of customizable KWS systems. The first category is a family of {\bf two-stage} approaches. These approaches usually employ an acoustic modeling stage followed by a non-trivial keyword search stage. {\em Graph search}, similar to the search process in automatic speech recognition (ASR), is widely used for customizable KWS in continuous speech because extending a well-trained ASR system to KWS tasks is natural and relatively straightforward. Here, an arbitrary keyword is represented as a short sequence of acoustic modeling units, such as monophones or graphemes. A filler model can be introduced to absorb non-keyword clues. Weighted finite state transducer (WFST) has been well accepted as the dominant graph search approach~\cite{hmm-filler-1, hmm-filler-2, hmm-filler-3, hmm-filler-4, hmm-filler-5}. How to build the decoding graph and utilize the scores of keyword and filler paths are the key issues to address in such systems. Later, a deep learning based acoustic model with a {\em Posterior handling} module is proposed for KWS~\cite{paper-guoguochen-dnn-kws}, simplifying the training and inference procedure. Following this, numerous improvements emerge, with some focusing on acoustic model architecture~\cite{cnn-kws,lstm-kws-1, lstm-kws-2,att-kws-1, att-kws-2} and others improving the posterior handling module~\cite{paper-dnn-kws-dp-automatic-gain,paper-decoding-1, paper-decoding-2}. Not all deep learning approaches are suitable for customizable KWS in continuous speech. A couple of approaches are proposed for customizable KWS. For example, ~\cite{06-openkws-e2e-1} proposes a transformer-based keyword localization method without requiring alignments for open-vocabulary KWS tasks, ~\cite{07-openkws-self} designs a detection module to formulate customizable KWS to a detection task. A common problem of the two-stage approaches is that the search process is resource-consuming and hence may limit their use in certain application scenes.

The second category of customizable KWS system is {\bf end-to-end} system based on representation matching. It involves two inputs: the enrolled keyword references and the speech data that needs to be detected. The fundamental idea is to ensure that the keyword speech representation is more similar to the enrolled reference pattern than the non-keyword speech. No complicated search method is required here, and the KWS decision can be made by simple representation feature matching, hence referred to as {\em end-to-end} approach. The classic system is query-by-example (QByE) using pre-enrolled audio samples~\cite{new-qbye-1,new-qbye-2,new-qbye-3,new-qbye-4, new-qbye-5,new-qbye-6}. Metric learning approaches have been proposed in recent years~\cite{icassp2023_qbye_01, icassp2023_qbye_02}. Another enrolled data type, which has become more popular these years, is the text format ~\cite{qbye-at-01,qbye-at-02-libriphrase, qbye-at-03-libriphrase-apple, qbye-at-04-libriphrase-korea}. The advantage of text-based enrollment is that it is user-friendly and not affected by the acoustic enrollment environment. However, with pure text enrollment, it is difficult to process speech with similar pronunciations and easy to falsely wake up when encountering confusing keywords speech~\cite{qbye-at-03-libriphrase-apple,qbye-at-04-libriphrase-korea}. Previous representation learning-based systems are mostly applied to pre-segmented isolated KWS. For customizable KWS in continuous speech, the streaming operation mode requires representation to be generated in consecutive sliding windows, which poses difficulty in effective negative sample generation. Furthermore, due to the co-articulation effect, representation learning is even more challenging. Apart from audio-text matching, confusing audio discrimination is also important. The neglect of audio confusion may cause a significant false alarm increase in continuous speech. To address these issues, in this paper, we propose a novel audio-text representation learning approach to achieve efficient and effective end-to-end KWS in continuous speech. Our core contributions can be summarized as follows:

\begin{figure*}[t]
\vspace{0em}
\centerline{\includegraphics[width=15cm]{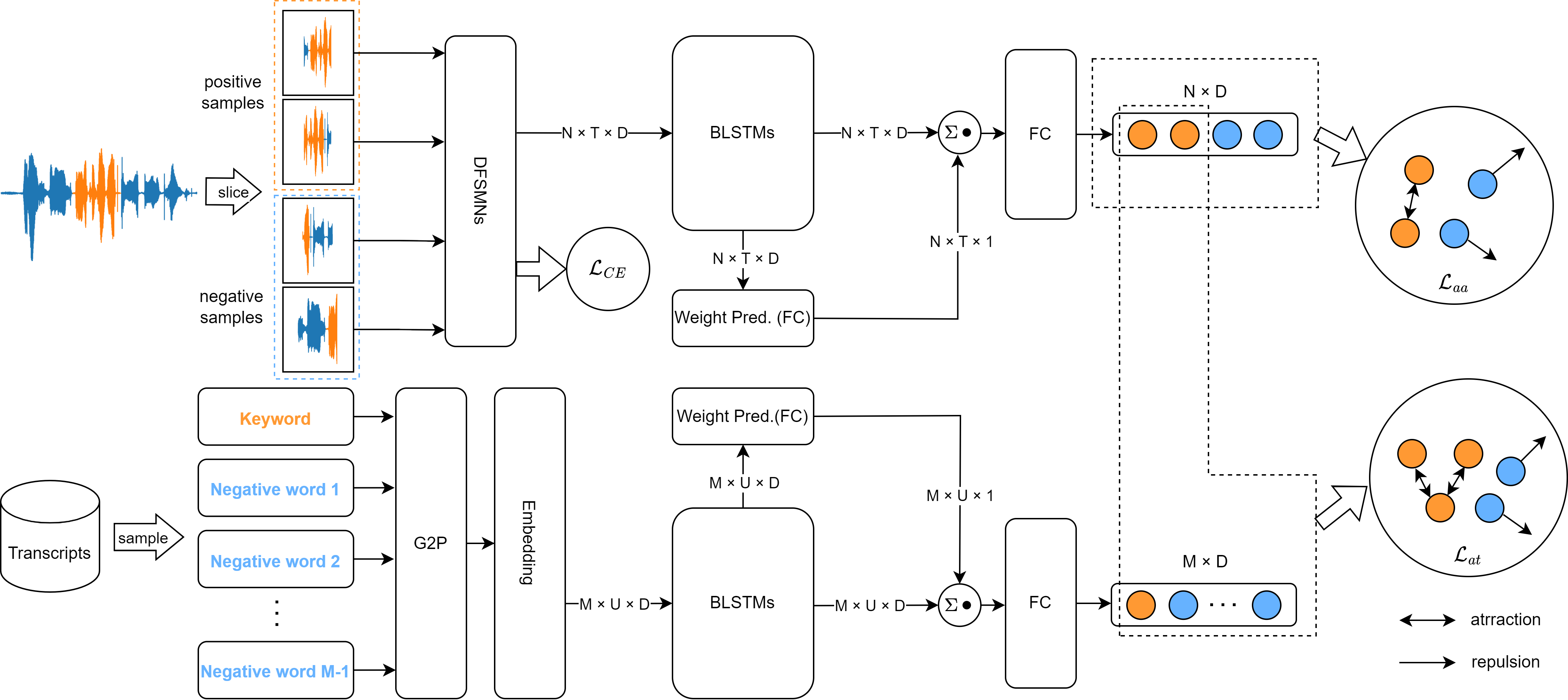}}
\caption{The overview of the whole framework.}
\label{fig:overview_figure}
\vspace{-1.5em}
\end{figure*}

\begin{itemize}
\setlength{\itemsep}{2pt}
\vspace{-5pt}
        \item In contrast to the previous audio-text matching strategy, we introduce a novel contrastive learning (CL) approach using audio discrimination (CLAD) to enhance the keyword representation quality. In addition, we design a data augmentation scheme to generate numerous similar and confusing segments. Experiments demonstrate that the additional audio discrimination strategy can significantly improve customizable KWS in continuous speech scenarios.  
        \item We propose to use sliding window level InfoNCE~\cite{CPC,simclr} for CLAD. Compared to other CL losses such as Triplet or SoftTriple, InfoNCE can accommodate rich positive and negative sample usage, which is particularly useful for effectively performing CL in continuous speech with both audio-text matching and audio-audio discrimination criteria. Note that, for isolated KWS, InfoNCE is shown to yield competitive performance compared to other classic CL losses.
        \item With representation from CLAD, we construct an end-to-end customizable KWS system. Compared to the two-stage approaches with the same acoustic model and various search strategies, our system can achieve both better performance and significant speed up. Also, the standard deviations among different keywords are lower, which shows its stability.
 \end{itemize}

\def\smallPlus{\texttt{+}}
\def\smallMinus{\texttt{-}}
\vspace{-8pt}
\section{End-to-end KWS Using Contrastive Learning with Audio Discrimination}
\label{sec:method}
In this section, we introduce the whole architecture and provide a detailed description of the training and inference procedure of our CLAD system.~\Cref{fig:overview_figure} presents the overview of the system.
\vspace{-5pt}
\subsection{The Frame-level AM Training}
\vspace{-5pt}
Firstly, a frame-level acoustic extractor is initially pre-trained to generate representations for the CLAD system. The modeling units in this context are the monophones, with forced alignments pre-processed to establish frame-level training targets. Following this, the acoustic model is trained based on cross-entropy (CE) criteria applied in a supervised learning manner. In our preliminary experiments, we discovered that the performance gap between freezing and unfreezing the Acoustic Model (AM) during the training of the CLAD system was marginal. Hence, we freeze the AM after pre-training, enabling it to supply only higher-level representations.

\subsection{Training Pairs and Training Strategy}
\label{sec:pairs}
For customizable Keyword Spotting (KWS) task, since the target keywords is not known during training, it is not possible to gather data for specific keywords on a scale comparable to the fixed KWS task. Consequently, effectively utilizing the general ASR data becomes a critical component for training a robust, customizable KWS model. We approach this by designing a simple data augmentation strategy, whereby sliding segments in speech near the keywords are used to generate a large volume of training pairs. Based on this, we develop a CLAD training strategy to learn the nuances of audio-text matching and audio-audio discrimination effectively.
 
\vspace{-7pt}
\subsubsection{Estimation of Sliding Window}
\vspace{-5pt}
To prevent mismatch between training and inference, we calculate the window length of an arbitrary keyword utilizing a consistent estimation function, which is defined as follows:
\begin{equation}
\label{equ:seg_len}
	L_{seg} = T_{mean} \cdot N_{phns} + L_{margin}\,.
\end{equation} 
Here, $N_{phns}$ denotes the number of the keyword phonemes, while $T_{mean}$ represents the average phoneme length calculated based on the training dataset. $L_{margin}$ is a hyper-parameter assigned for providing an additional length to the segment.

We add $L_{margin}$ to ~\Cref{equ:seg_len} due to two key reasons: (1) To accommodate keywords exceeding the average pronunciation duration. (2) To encapsulate contextual information proximate to the keyword, which aids in considering the co-articulation effect and discriminates confused content during model training. In our experiments, $T_{mean}$ is estimated based on LibriSpeech dataset and set at 90ms. Simultaneously, we set $L_{margin}$ at 300ms to accommodate approximately three or four additional phonetic units beyond the keyword phoneme sequence.




\vspace{-7pt}
\subsubsection{Contrastive Learning for Audio-text Matching}           \vspace{-5pt}     
\label{sec:audio-text-pairs}
To effectively learn the matching correlation between textual and audio representations, it is essential to construct a substantial quantity of text-audio pairings during the training process. As depicted in~\Cref{fig:overview_figure}, we randomly sample arbitrary words regarded as keywords for the given utterance in this round. In each mini-batch, we denote the j-th sampled textual keyword in the i-th utterance as $W_{i,j}$, all N corresponding positive audio segments as $A_{i,j}^{p1}, A_{i,j}^{p2},  \ldots,A_{i,j}^{pN}$, and all M corresponding negative audio segments as $A_{i,j}^{n1}, A_{i,j}^{n2},\ldots, A_{i,j}^{nM}$. The process of constructing positive segment $A_{i,j}^{px}$ and negative segment $A_{i,j}^{nM}$ for each keyword is discussed in \Cref{sec:audio-pairs}.

When constructing audio-text pairs, we solely consider positive audio segments $A_{i,j}^{pk}$ and the corresponding keywords $W_{i,j}$, omitting negative audio segments $A_{i,j}^{nx}$. Each positive audio-text pair is composed of an audio segment $A_{i,j}^{pk}$ and the corresponding keyword $W_{i,j}$. Positive audio samples $A_{i,j}^{pk}$ and all textual keywords $W$ within the mini-batch (excluding the correlating keyword $W_{i,j}$) are considered negative audio-text pairings. For a given mini-batch, the loss for audio-text pairs is defined as follows:

\begin{equation}
\label{equ:at}
\resizebox{0.91\hsize}{!}{
  $\mathcal{L}_{at} = - \sum\limits_{i,j,k} \log \frac
  {\exp \left(\operatorname{Sim} \left( \boldsymbol{A_{i,j}^{pk}}, \boldsymbol{W_{i,j}} \right) / \tau_{at} \right)}
  {\exp \left(\operatorname{Sim} \left( \boldsymbol{A_{i,j}^{pk}}, \boldsymbol{W_{i,j}} \right) / \tau_{at} \right)
  + \sum\limits_{w\neq(i,j)} \exp \left(\operatorname{Sim} \left( \boldsymbol{A_{i,j}^{pk}}, \boldsymbol{W_{w}} \right) /\tau_{at}\right)
  }
  \,,$	
}	
\end{equation}
where $at$ denotes the abbreviation of audio-text pairs. $\operatorname{Sim}$ represents a function to ascertain the degree of similarity, while $\tau$ signifies a temperature hyper-parameter for controlling the concentration of features in the representation space.

\vspace{-7pt}
\subsubsection{Contrastive Learning for Audio-audio Discrimination}
\vspace{-5pt}
\label{sec:audio-pairs}
In continuous streaming scenarios, simply considering matching text representation with audio representation is insufficient. The issue arises from the fact that during the streaming sliding window inference process, numerous counter example segments have not been encountered during the training phase. For instance, certain fragments might contain portions of the preceding word in addition to parts of the wake word. Contrastive learning (CL) with audio-text struggles to distinguish such challenging negative audio examples on its own. To address this issue, we propose CL with audio discrimination (CLAD).

We slice positive and negative audio segments from the original audio for each keyword based on the ratio of overlaps between the estimated window and the ground-truth keyword location, Any two $A_{i,j}^{pk}$ and $A_{i,j}^{pl}$ ($k \neq l$) are considered as a positive pair. Any $A_{i,j}^{pk}$  and $A_{i,j}^{nx}$  are considered as a negative pair. As there are $C_{N}^{2}$ positive pairs for each sampled keyword, the number of audio pairs is enormous. For a mini-batch, the loss of audio-audio pairs is defined as follows:

\begin{equation}
\label{equ:aa}
\resizebox{0.91\hsize}{!}{
	$\mathcal{L}_{aa} = - \sum\limits_{i,j}\sum\limits_{\;k,l,k\neq l} \log \frac
	{\exp \left(\operatorname{Sim} \left( \boldsymbol{A_{i,j}^{pk}}, \boldsymbol{A_{i,j}^{pl}} \right) / \tau_{aa}\right)}
	{\exp \left(\operatorname{Sim} \left( \boldsymbol{A_{i,j}^{pk}}, \boldsymbol{A_{i,j}^{pl}} \right) / \tau_{aa}\right) 
	+\sum\limits_{x=1}^{M} \exp \left(\operatorname{Sim} \left( \boldsymbol{A_{i,j}^{pk}}, \boldsymbol{A_{i,j}^{nx}} \right) /\tau_{aa}\right)}
	\,,$
}
\end{equation}
where $aa$ is the abbreviation of audio-audio pairs. $\operatorname{Sim}$ and $\tau$ are the same as mentioned before.

The final CLAD representation loss criteria is defined as follows:
\begin{equation}
\label{equ:loss}
	\mathcal{L} = \alpha  \mathcal{L}_{aa} +  \mathcal{L}_{at} \,.
\end{equation}
Here, $\alpha$ is a hyper-parameter to balance the ratio of matching and discriminating parts.

Empirically, we find the temperature $\tau_{at} = 0.12$ in~\Cref{equ:at}, the temperature $\tau_{aa} = 0.2$ in~\Cref{equ:aa}, and $\alpha = 0.15$ in~\Cref{equ:loss} work well in our experiments. In the following discussions, unless otherwise specified, these three hyper-parameters are set to the specified values.

\vspace{-7pt}
\subsection{End-to-end KWS with Keyword Representation Matching}

The confidence level in keyword existence is assessed through the similarity between the embeddings of the audio segment and the enrollment textual keyword, like other embedding-based systems~\cite{qbye-at-01,qbye-at-02-libriphrase}. However, our operations are conducted on a continuous speech stream rather than isolated segments. We begin by estimating the audio segment length by ~\Cref{equ:seg_len} and segmenting the test audio track into slices, ensuring half overlap between adjacent segments. Following this, we forward the audio segments to compute the similarity with the keyword representation consecutively. The highest similarity score is acknowledged as the keyword score. If the keyword score surpasses the threshold, we reset the model state and introduce a 1-second cooldown period to prevent recurrent wake-up on the same keyword.



\vspace{-7pt}
\section{experiment configuration}
\label{sec:exp}

\vspace{-3pt}
\subsection{Datasets}
\vspace{-3pt}
\label{sec:exp_data}

We evaluate our CLAD system in two scenarios: (1) Continuous KWS, where the keyword can be situated anywhere within a given sentence. To perform this evaluation, we construct \textbf{the kws-version of LibriSpeech}. (2) Isolated KWS, where each sample solely comprises either a keyword or a non-keyword. As previous representation learning-based methods were not designed for a continuous mode, we compare our approach with others using \textbf{LibriPhrase}. The following are the details of these datasets:

\begin{itemize}
\setlength{\itemsep}{2pt}
        \item \textbf{The kws-version of LibriSpeech}. We utilize the full 960 hours of LibriSpeech as training data. Following the strategy in~\cite{paper-decoding-1}, we choose the 5 to 50 most frequent words that contain at least six phonemes each for test-clean and test-other, respectively. The entirety of the audio files, barring those that contain the keywords, are designated to the corresponding false alarm dataset, each equating to approximately 3 hours.

	\item \textbf{LibriPhrase}~\cite{qbye-at-02-libriphrase}. LibriPhrase is also derived from LibriSpeech, which is constructed to evaluate isolated KWS for audio-text representation learning-based methods. The training dataset is constructed from train-clean-460, while the test dataset is constructed from train-others-500. LibriPhrase test dataset is divided into two parts: easy negatives $\textbf{\text{LP}}_\textbf{\text{E}}$ and hard negatives $\textbf{\text{LP}}_\textbf{\text{H}}$. Each episode in $\textbf{\text{LP}}_\textbf{\text{E}}$  and $\textbf{\text{LP}}_\textbf{\text{H}}$  contains 3 positive and 3 negative audio-text pairs. Detailed description can be found in \cite{qbye-at-02-libriphrase}. For a fair comparison, we re-train our CLAD system in 460h train-clean data, while using $\textbf{\text{LP}}_\textbf{\text{E}}$ and $\textbf{\text{LP}}_\textbf{\text{H}}$ to evaluate.

 \end{itemize}

\vspace{-7pt}
\subsection{Implementation details}
\vspace{-3pt}
The acoustic model utilized in all experiments comprises 5 layers of Deep Feedforward Sequential Memory (DFSMN)~\cite{DFSMN}, with a hidden size of 512 and a projection size of 128. The left context of the DFSMN module is set to 10, while the right context is set to 1. Both the acoustic encoder and textual encoder consist of three 128-dimensional BLSTMs with a 64-dimensional projection. The output of the last layer of BLSTMs is weighted-summed by the output of the weight prediction layer, which maps the 128-dimension vector to a scalar. The acoustic and semantic weighted-summed vectors are then individually mapped by two distinct 128-dimensional fully connected (FC) layers to the contrastive embedding space. We apply cosine similarity to obtain the final similarity score of the two contrastive embeddings. The total number of parameters in our system is approximately 2.2 million, which is suitable for on-device KWS tasks. We utilize the SGD~\cite{SGD} optimizer with an initial learning rate of 5e-6. The mini-batch size is set to 12,288 frames. If the loss on the validation set does not decrease, the learning rate is halved. Furthermore, if the loss on the validation set does not decrease for three consecutive rounds, the training process is finished.

Two widely used two-stage streaming KWS baselines are compared with our end-to-end KWS model. The first is a well-known graph-based approach containing an acoustic model and keyword/filler decoding graphs. The second system is the most widely used system in the DNN-KWS, which contains an acoustic model and a post-processing module. The acoustic models used in two baselines and our CLAD are the same.



\vspace{-7pt}
\section{results and analysis}
\label{sec:rec_and_ana}

\vspace{-3pt}
\subsection{Performance on isolated KWS}
\vspace{-3pt}
In this part, we compare the performance of CLAD with other representation learning methods for isolated KWS, as shown in~\Cref{table:table1}. The results of CMCD and CLAD show that even though our approach is designed for continuous speech streams, we can achieve comparable or even slightly better performance compared to the mainstreaming method in isolated KWS. The results in rows 2, 3, and 4 show that InfoNCE is well suited for the customizable KWS task based on representation learning, and it achieves better performance than the other two classical contrastive learning losses. Comparison of lines 4 and 5 shows that even in the context-free environment of isolated keywords, our proposed CLAD strategy still achieves some performance gains.

\begin{table}[t]
  \centering
  \begin{resizebox}{0.75\columnwidth}{!}
  {
    \begin{tabular}{ c c c c c c }
     \toprule
     \multirow{2}*{\textbf{Model}} & \multirow{2}*{\textbf{CL}} & \multicolumn{2}{c}{\textbf{EER}(\%)} & \multicolumn{2}{c}{\textbf{AUC}(\%)} \\
     \cmidrule(lr){3-4} \cmidrule(lr){5-6}
     ~ & ~ & $\textbf{\text{LP}}_\textbf{\text{E}}$ & $\textbf{\text{LP}}_\textbf{\text{H}}$ & $\textbf{\text{LP}}_\textbf{\text{E}}$ & $\textbf{\text{LP}}_\textbf{\text{H}}$ \\
     \midrule 
     CMCD~\cite{qbye-at-02-libriphrase}& \ding{56} & \textbf{8.42} & 32.90 & 96.70 & 73.58 \\
     \midrule
     Triplet~\cite{02-openkws-embed-1}& \ding{51} & 32.75  & 44.36 & 63.53 & 54.88 \\
     SoftTriple~\cite{new-qbye-5}& \ding{51} & 28.74 & 41.95 & 78.74 & 62.65 \\
     \midrule
     InfoNCE & \ding{51} & 8.99 & 32.51 & 96.85 & 74.87 \\
     CLAD & \ding{51} &8.65 & \textbf{30.30} & \textbf{97.03} & \textbf{76.15} \\
    \bottomrule
    \end{tabular}%
   }
   \end{resizebox}
   \linespread{0.9}
   \caption{Comparison of EER and AUC score in $\textbf{\text{LP}}_\textbf{\text{E}}$ and $\textbf{\text{LP}}_\textbf{\text{H}}$. CL denotes whether the method is based on contrastive learning.}
  \label{table:table1}
\vspace{-12pt}
\end{table}

\vspace{-7pt}
\begin{figure}[b]
\vspace{-0.36cm}
    \centering
    \includegraphics[width=0.9\linewidth]{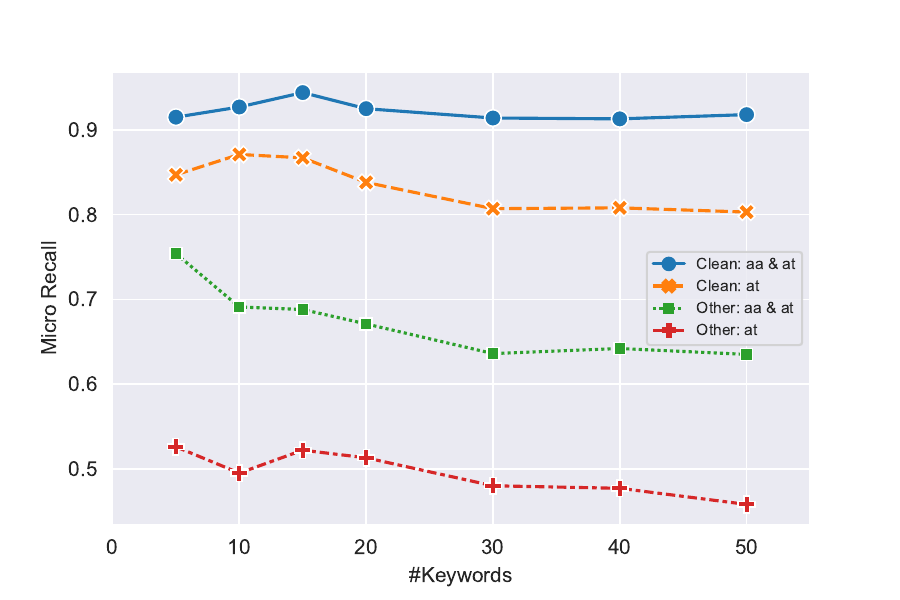}

    \vspace{-0.3cm}
    \linespread{0.9} 
    \caption{Results of models trained by audio-audio and audio-text pairs or only audio-text pairs. ``Clean"  means the results of the test-clean dataset, and ``Other" means the results of the test-other dataset. ``aa" denotes CLAD, while ``at" denotes CL without AD.}

    \label{fig:f1_score}
\vspace{-0.5cm}
\end{figure}

\subsection{Performance on continuous KWS}
\vspace{-3pt}
We present the experiment results in~\Cref{table:res}. The performance of the graph-based baseline is superior on high-frequency keywords when the test dataset is simple, while the performance decreases dramatically in complex acoustic environments or on relatively rare keywords. Compared with the two baseline models, except the graph-based results of the relatively small number of keywords (5, 10, 15) in the test-clean dataset, our model can yield consistent gains from 5 to 50 words on both the simple test-clean dataset and the challenging test-other dataset.

When the number of keywords is 5 or 10, the results reflect more about the ability to model high-frequency words. In contrast, when the number of keywords is relatively large, like 40 or 50, the results reflect more on the generalization of the model to arbitrarily specify keywords. For the customizable KWS system, these two aspects are equally crucial for researchers or users. Good results on numerous keywords provide the ability to select an arbitrary keyword for users, and good results on high-frequency words offer the possibility of further customization after the keywords are determined.

\begin{table}[t]
  \centering
  \begin{resizebox}{1.0\columnwidth}{!}
  {
    \begin{tabular}{ c  c c c c c c c c }
      \toprule
      \multirow{2}*{\textbf{Dataset}} & \multirow{2}*{\textbf{Model Type}} &  \multicolumn{7}{c}{\textbf{\#Keywords}} \\
      \cmidrule(lr){3-9}
      																						~ & ~ & 5 & 10 & 15 & 20& 30 & 40& 50\\
      \midrule
      \multirow{3}*{test-clean} & Graph-based &    \textbf{0.943} & \textbf{0.948} &  \textbf{0.958} & 0.916 & 0.901 &  0.899 & 0.907 \\ 
        ~ & Post. Handl. &  0.881 & 0.906 &  0.923 & 0.887 & 0.876 &  0.876 & 0.872 \\
      ~ &  CLAD   & 0.915 & 0.927 &  0.944  & \textbf{0.925} & \textbf{0.914} &  \textbf{0.913} & \textbf{0.918} \\ \midrule
      
    \multirow{3}*{test-other} &  Graph-based  &  0.696 & 0.655 & 0.657 & 0.660 & 0.625 & 0.617 & 0.605 \\
     ~ & Post. Handl. &   0.678 & 0.636 &  0.654  & 0.643 & 0.614 &  0.623 & 0.612  \\
     ~  & CLAD &  \textbf{0.754} & \textbf{0.691} &  \textbf{0.688}  & \textbf{0.671} & \textbf{0.636} &  \textbf{0.642} & \textbf{0.635}  \\ \bottomrule

    \end{tabular}%
   }
   \end{resizebox}
   \linespread{0.9}
   \caption{Results of the baseline models and the proposed model. We present micro recalls for 5, 10, 15, 20, 30, 40, and 50 keywords, respectively, under the condition that the number of false alarms is two on the corresponding false alarm dataset.}
   \label{table:res}
   \vspace{-12pt}
\end{table}

\vspace{-7pt}
\subsection{The importance of audio discrimination}
\vspace{-3pt}
To explore the effectiveness of the proposed CLAD, we conduct ablation experiments using audio-text CL without audio discrimination and show the results in ~\Cref{fig:f1_score}. The performance degradation is dramatic. It is not surprising that audio-audio pairs are essential for improving the model performance, as they introduce discriminative information to distinguish locally similar negative samples with keywords in continuous speech, significantly improving performance.

\vspace{-7pt}
\subsection{Inference Speed}
\vspace{-3pt}
In this part, we present the relative speed acceleration (RSA) to explore the inference latency. RSA is defined as the ratio of the execution time of the benchmark and the model. As shown in~\Cref{table:rtf}, the proposed CLAD system has a breakneck inference speed compared with other baselines. This can be primarily attributed to our methodology, which employs an end-to-end representation vector similarity matching approach. This effectively eliminates the need for a time-consuming search module.
\begin{table}[h]
  \centering
  {
    \begin{tabular}{ l |c |c | c }
      \toprule
      model & Graph-based & Post. Handl. & CLAD \\
     \midrule
      RSA & 1.00X & 0.38X & \textbf{16.95X}\\ 
	 \bottomrule
    \end{tabular}%
   }
    \linespread{0.9}
   \caption{Comparison of RSA of the baselines and CLAD. A higher RSA indicates a faster inference speed.}
  \label{table:rtf}
  \vspace{-9pt}
\end{table}

\vspace{-0.4cm}
\section{conclusions}

In this paper, CLAD, a contrastive learning (CL) based method with audio discrimination, is proposed for representation learning in KWS in continuous speech. In contrast to previous CL approaches which all focus on audio-text matching, we further introduce audio discrimination criterion for robust representation learning. With CLAD, we propose an efficient end-to-end KWS for customizable KWS in continuous speech. Experiments show that, although CLAD is proposed for continuous speech, it can achieve comparable or superior performance compared with other audio-text representation matching methods on KWS in isolated speech. When comparing with the widely used two-stage approaches on KWS in continuous speech, our method got best performance in most testing configurations. In addition, significant speed up is achieved compared to two-stage approaches. 


\newpage
\linespread{0.9}
\bibliographystyle{styles/IEEEbib}
\bibliography{citations/refs}
\end{document}